\documentclass{PoS}

\PoS{PoS(HEP2005)281}

\title{The Angular Distribution of $B^0 \rightarrow K^{*0}(\rightarrow K^- \pi^+)l^+l^-$ at Large Recoil in and Beyond the SM}

\ShortTitle{The Angular Distribution of $B^0 \rightarrow
K^{*0}(\rightarrow K \pi)l^+l^-$ in and beyond the SM}

\author{\speaker{Joaquim Matias}
\\
        Departament de F\'{\i}sica, Universitat Aut\`onoma de Barcelona, Spain\\
        E-mail: \email{matias@ifae.es}}

\abstract{ We discuss, in detail, the $K^*$ polarization states in
the exclusive $B$ meson decay ${B}^0 \to K^{*0}(\to K^- \pi
^+)l^+l^-$ ($l=e, \mu, \tau$) in the low dilepton mass region. We
focus on the study of the angular distribution of this decay that
provides valuable information on the $K^*$ spin amplitudes $\app$,
$\ap$, $\al$. This can give us a handle on non-standard
interactions that cannot be proved through measurements of the
branching ratio and lepton forward-backward asymmetry. We explore
the transverse asymmetries
$A^{(1)}_{\trans}({s})$, $A^{(2)}_{\trans}({s})$, $K^*$ polarization
parameter
$\alpha_{K^*}(s)$, the fraction of
$K^*$ polarization $F_L(s)$ and $F_T(s)$ and the corresponding integrated
observables
 at NLL order, including factorizable and
non-factorizable corrections. We find, in particular, that the
dependence on hadronic uncertainties for the transverse
asymmetries turns out to be
very small. This allow us to distinguish which
observables are better suited to look for physics beyond the SM.
Finally, we study in a model independent way the implications of
New Physics for these observables.}

\FullConference{International Europhysics Conference on High Energy Physics\\
         July 21st - 27th 2005\\
         Lisboa, Portugal}
\def \be{\begin{equation}}
\def \ee{\end{equation}}
\def \bea{\begin{eqnarray}}
\def \eea{\end{eqnarray}}
\def \ben{\begin{enumerate}}
\def \een{\end{enumerate}}
\def \bit{\begin{itemize}}
\def \eit{\end{itemize}}
\def \Bbar{\overline{\kern -0.24em B}}
\def \al{A_0}
\def \ap{A_{\|}}
\def \app{{A}_{\bot}}
\def \kstar{{K^*}}
\def \trans{{T}}
\def \bm{\boldmath}

\def \braket#1#2#3{\langle #1|#2| #3\rangle}

\def \cseff{C_7^{\rm {eff}}}
\def \cseffP{{C_7^{\rm {eff}}}^\prime}
\def \ceff{C_9^{\rm {eff}}}

\def \cten{C_{10}}

\def \a{\alpha}

\def \g{\gamma}
\def \G{\Gamma}

\def \m{\mu}
\def\euro#1#2#3{{Eur. Phys. J. C} {\bf #1}, #3 (#2)}

\def\ibid#1#2#3{{\it ibid.\/}~{\bf#1}, #3 (#2)}
\def\ib#1#2#3{{\bf#1}, #3 (#2)}
\def\jhep#1#2#3{{J.~High Energy Phys.} {\bf #1}, #3 (#2)}

\def\np#1#2#3{{Nucl.~Phys.}~{\bf B#1}, #3 (#2)}

\def\pl#1#2#3{{Phys.~Lett. B}~{\bf #1}, #3 (#2)}

\def\prd#1#2#3{{Phys.~Rev. D}~{\bf #1}, #3 (#2)}

\begin{document}
We have recently entered into a Precision Flavour Physics Era
thanks to the huge experimental effort in present $B$-facilities
and the improvement in our  theoretical knowledge of certain $B$
meson decays. This has changed our way of looking at New Physics
(NP). We are moving from the simple yes/no question concerning the
presence of NP to the {\it measurement} of NP parameters
\cite{london}. In particular, we are starting to design
observables mostly sensitive to a specific type of NP (isospin
breaking, right handed currents, etc.). The observables that we
will discuss here, based on the decay $B^0 \rightarrow K^*
(\rightarrow K \pi)l^+ l^-$, are an example of this approach. They
are built to test the chiral structure of the fundamental theory
that is beyond the SM, in particular, the presence of right-handed
currents \cite{melikhov,our}. The decay $B^0 \rightarrow K^*
(\rightarrow K \pi)l^+ l^-$  provides valuable information in many
different ways, by means of the branching ratio \cite{ali}, the
lepton forward-backward asymmetry \cite{feldmann} and, finally,
the angular distribution of this decay \cite{our,kruger}. It has
been shown that the study of the angular distribution of the
4-body final state provides information on the $K^*$ spin
amplitudes that are useful to search for right-handed currents
\cite{melikhov,our}. Our goal \cite{our} will be to identify which
are the most robust observables to search for this type of NP.

The effective hamiltonian for decays governed by the quark
transition  $b\rightarrow s l^+l^-$ is given by $\label{heff}
{\mathcal{H}}_{\rm eff}=-\frac{4 G_F}{\sqrt{2}} V_{tb}^{} V_{ts}^*
\sum_{i=1}^{10} [C_i (\mu) {\mathcal{O}}_{i}(\mu)  + C_i^\prime
(\mu) {\mathcal{O}}_{i}^\prime (\mu)]$,  where we have added
together with the SM operators the chirally flipped partners to
describe right-handed currents. Our main focus here will be the
magnetic penguins and semi-leptonic operators: ${\mathcal{O}}_{7}
= \frac{e}{16\pi^2} m_b (\bar{s} \sigma_{\mu \nu} P_R b) F^{\mu
\nu}$, ${\mathcal{O}}_{9} = \frac{e^2}{16\pi^2} (\bar{s}
\gamma_{\mu} P_L b)(\bar{l} \gamma^\mu l)$ and
${\mathcal{O}}_{10}=\frac{e^2}{16\pi^2} (\bar{s}  \gamma_{\mu} P_L
b)(  \bar{l} \gamma^\mu \gamma_5 l)$ together with the chirally
flipped operator  ${\mathcal{O}}_{7}^\prime = \frac{e}{16\pi^2}
m_b (\bar{s} \sigma_{\mu \nu} P_L b) F^{\mu \nu}$. We will also
discuss how to include  the chiral partners of the operators
${\mathcal{O}}_{9}, {\mathcal{O}}_{10}$ although they will not be
considered in the analysis.

Given the effective Hamiltonian \ref{heff} one can compute the
matrix element: \begin{eqnarray}\label{matrix:ele} {\mathcal M}
&=& \frac{G_F\a}{\sqrt{2}\pi}V_{tb}^{}V_{ts}^*\bigg\{
\bigg[\ceff\braket{K \pi}{(\bar{s}\g^{\mu}P_Lb)}{B}
-\frac{2m_b}{q^2}
\braket{K\pi}{\bar{s}i\sigma^{\mu\nu}q_{\nu}(\cseff P_R+\cseffP
P_L)b}{B}\bigg]
(\bar{l}\g_{\m}l)\nonumber\\
&+&
\cten\braket{K\pi}{(\bar{s}\g^{\mu}P_Lb)}{B}(\bar{l}\g_{\mu}\g_5
l)\bigg \}, \end{eqnarray} where $q$ is the four-momentum of the
lepton pair and $m_b(\mu)$ is the running mass in the MS scheme.
The hadronic matrix elements entering Eq.(\ref{matrix:ele}) are
parameterized by means of a narrow width approximation
\cite{kruger} in terms of seven $B \rightarrow K^*$ form factors
(see \cite{our,ali}). The crucial point is how to deal with the
heavy to light form factors. Either, one can take the approach of
QCD sum rules\cite{sumrules} and try to compute them. Or following
the key observation that in the limit where the initial hadron is
heavy and the final meson has a large energy it is possible to
reduce the $A_i(s), T_i(s)$ and $V(s)$ form factors to only two
universal form factors ($\xi_\perp$ and $\xi_{||}$)\cite{charles}.
However, these relations are valid for the soft contribution to
the form factors at large recoil, i.e, we are restricted to the
kinematical region
 where $ E_\kstar$ is
large and the dilepton mass is small. Moreover, these relations
are violated by symmetry breaking corrections of order $\alpha_s$
and $1/m_b$. We will include here the factorizable and
non-factorizable corrections in $\alpha_s$ \cite{beneke}. There
are also possible quark-antiquark resonant intermediate states
contributions as well as other long distance effects that will be
presented elsewhere\cite{jm}.

Assuming the $K^*$ to be on the mass shell, the differential decay
rate of the decay $B^0 \rightarrow K^* (\rightarrow K \pi)l^+ l^-$
is described in terms of four independent kinematical variables:
the lepton-pair invariant mass, $s$, and the three angles
$\theta_l$, $\theta_{K^*}$, $\phi$. In terms of these variables,
the differential decay rate can be written as \cite{kruger} $
\frac{d^4\G}{ds\,d\cos\theta_l\, d\cos\theta_{K^*}\, d\phi} =
\frac{9}{32 \pi}\sum_{i=1}^9
I_i(s,\theta_{K^*})f_i(\theta_l,\phi), $ within the physical
region of phase space. $I_i$ depend on products of the four $K^*$
spin amplitudes $\app$, $\ap$, $\al$, $A_t$, and $f_i$ are the
corresponding angular distribution functions. Given the matrix
element Eq.(\ref{matrix:ele}) one can compute the transversity
amplitudes (see \cite{our} for $\al$, $A_t$) at LO that reads:
\be\label{a_perp} A_{\bot L,R}=N \sqrt{2} \lambda^{1/2}\bigg[
(\ceff\mp\cten)\frac{V(s)}{m_B +m_\kstar}+\frac{2m_b}{s} (\cseff +
\cseffP) T_1(s)\bigg], \ee \be\label{a_par} A_{\| L,R}= - N
\sqrt{2}(m_B^2- m_\kstar^2)\bigg[(\ceff\mp \cten) \frac{A_1
(s)}{m_B-m_\kstar} +\frac{2 m_b}{s} (\cseff - \cseffP)
T_2(s)\bigg]. \ee In the SM (in particular $\cseffP=0$) we recover
the naive quark-model prediction $A_{\bot}=-A_{\|}$ in the LEET
limit. 
Notice also that, in this limit, the transverse (longitudinal and
time-like) $K^*$ polarizations involve only $\xi_\perp$($\xi_\|$).
Finally, the chirality-flipped operators
${\mathcal{O}}_{9,10}^\prime$ can be included in the above
amplitudes by the replacements $C^{(\mathrm{eff})}_{9,10}\to
C^{(\mathrm{eff})}_{9,10}+ C^{(\mathrm{eff})\prime}_{9,10}$ in
$A_{\perp L,R}$ and $C_{9,10}^{(\mathrm{eff})}\to
C^{(\mathrm{eff})}_{9,10}- C^{(\mathrm{eff})\prime}_{9,10}$ in
$A_{\| L,R}$.

\section{Observables}
We consider a series of observables constructed on the basis of
the $K^*$ spin amplitudes, whose magnitude and relative phases are
obtained from the study of the angular distribution of this decay
\cite{our}. In order to minimize the uncertainties due to hadronic
form factors we consider observables that involve ratios of
amplitudes. Using $ A_i A^*_j\equiv A^{}_{i L}(s) A^*_{jL}(s)+
A^{}_{iR}(s) A^*_{jR}(s) \quad (i,j  = 0, \|, \perp)$ we
investigate the following observables: (i) Transverse asymmetries
$A^{(1)}_{\trans}({s})=\frac{-2 {\rm Re}(\ap^{}\app^*)}{|\app|^2 +
|\ap|^2}$, $A^{(2)}_{\trans}({s})=\frac{|\app|^2 -
|\ap|^2}{|\app|^2 + |\ap|^2}. $  (ii) $K^*$ polarization parameter
$\label{def:pol:param} \alpha_{K^*}(s) = \frac{2|{
A}_0|^2}{|{A}_{\|}|^2 + |{A}_{\perp}|^2}-1. $ (iii) Fraction of
$K^*$ polarization $\label{def:frac:pol} F_L(s) =
\frac{|{A}_0|^2}{|{ A}_0|^2 + |{A}_{\|}|^2 + |A_\perp|^2}$ and
$F_T(s) = \frac{|{A}_\bot|^2+|{A}_\||^2}{|{ A}_0|^2 + |{A}_{\|}|^2
+ |A_\perp|^2}$,  so that $\alpha_{K^*}=2F_L/F_T-1$. (iv)
Integrated quantities ${\mathcal A}^{(1)}_{\trans}$, ${\mathcal
A}^{(2)}_{\trans}$, $\mbox{\bm $\alpha$}_{K^*}$, and ${\mathcal
F}_{L,T}$, which are obtained from (i) to (iii) by integrating
numerator and denominator separately over the dilepton invariant
mass.
\section{SM predictions for the observables and New Physics Impact}
We included factorizable and non-factorizable corrections at NLL
order by replacing  $\cseff T_i \rightarrow {\cal T}_i$, and
$\ceff \rightarrow C_9\quad (i=1,2,3)$ in the transversity
amplitudes \cite{our} and taking  $C_{9,10}$  at NNLL order (
${\cal T}_i$ are defined in \cite{beneke}). We are interested here
in exploring the impact of NLL corrections on the observables and
the sensitivity to the variation of the theoretical parameters.
The main sources of uncertainties are the dependence on soft form
factors $\xi_{\perp,\|}(0)$, the scale dependence, the ratio
$m_c/m_b$ that mainly affects the matrix elements of the
chromomagnetic operator, and the error associated to the rest of
input parameters (masses and decay constants) which are taken in
quadrature.

The results are  that $A^{(1,2)}_{\trans}({s})$ are the most
promising observables \cite{our}: the impact of NLL corrections
are negligible including uncertainties (see Fig 1). On the
contrary, $\alpha_{K^*}(s)$ receives a strong impact of the NLL
correction and a wide error band mainly due to the poorly known
$\xi_\perp(0)$ form factor. Concerning $F_L$ and $F_T$ the fact
that $A_0$ enters as a normalization factor moderates slightly the
influence of $\xi_\perp(0)$. We have also computed the SM
predictions for the integrated observables over the low dimuon
mass region $2 m_\mu \leq M_{\mu+\mu-} \leq 2.5 {\rm GeV}$ at NLL:
${\mathcal A}^{(1)}_{\trans}=0.9986\pm 0.0002$, ${\mathcal
A}^{(2)}_{\trans}=-0.043\pm 0.003$, $\mbox{\bm
$\alpha$}_{K^*}=3.47\pm 0.71$, ${\mathcal F}_{L}=0.69\pm 0.03$ and
${\mathcal F}_{T}=0.31\pm 0.03$.

Finally, we performed a model independent analysis \cite{our} of
the implications of right-handed currents for these observables.
Since the low dimuon mass region is dominated by the photon pole
$\cseffP/s$, the contribution from chirality flipped operators
${\cal O}_{9,10}^\prime$ will be subdominant. We imposed also the
constrain coming from $BR(B\rightarrow X_s \gamma)$. The results
were that even a small contribution from RH currents in $\cseffP$
produces a striking effect in both asymmetries ${\mathcal
A}^{(1)}_{\trans}$ and ${\mathcal A}^{(2)}_{\trans}$ (see Fig.1).
Moreover, the latter is sensitive also to the sign of $\cseffP$.
Interestingly, even if we allow for the presence of NP in ${\cal
O}_{9,10}$ up to 20\% still it is possible to use these
asymmetries to determine the magnitude and sign of the
contribution of RH currents into ${\cal O}_7^\prime$. On the
contrary, $F_{L,T}$ and specially $\alpha_{K*}$ are not
particularly suitable to disentangle this type of NP effects from
hadronic uncertainties due to their strong dependence on the
poorly unknown parameter $\xi_\perp(0)$. In conclusion, we have
shown that ${\mathcal A}^{(1,2)}_{\trans}$ are a useful probe of
the electromagnetic penguin operator ${\cal O}_7^\prime$.\\ {\it
Acknowledgements}: I acknowledge financial support from the Ramon
y Cajal Program, FPA2002-00748 and PNL2005-41.

%
\begin{figure}
\begin{center}
\hspace*{-0.1cm}\includegraphics[scale=0.46]{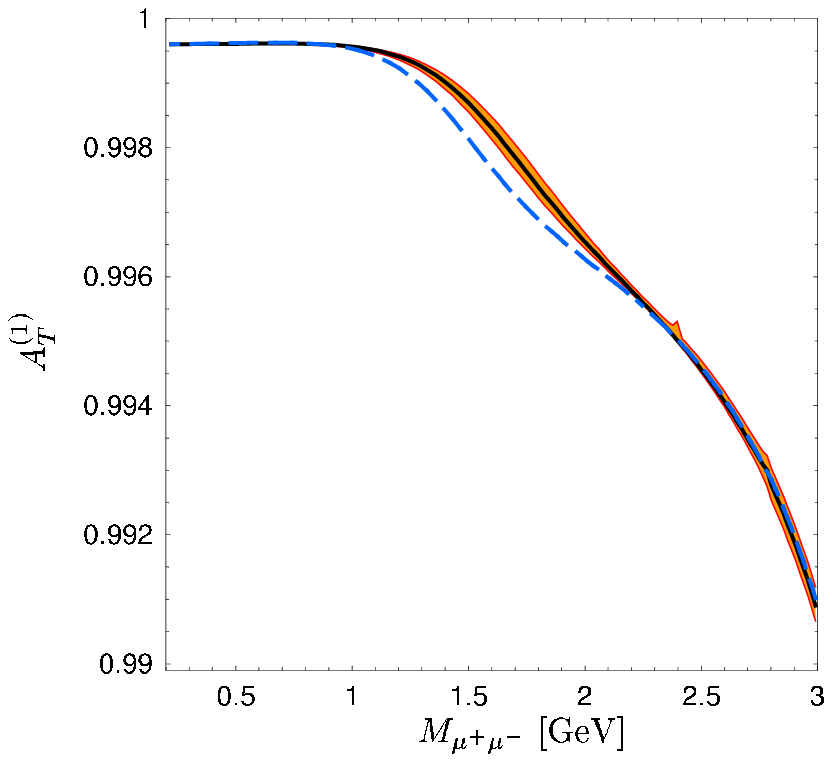}
\includegraphics[scale=0.46]{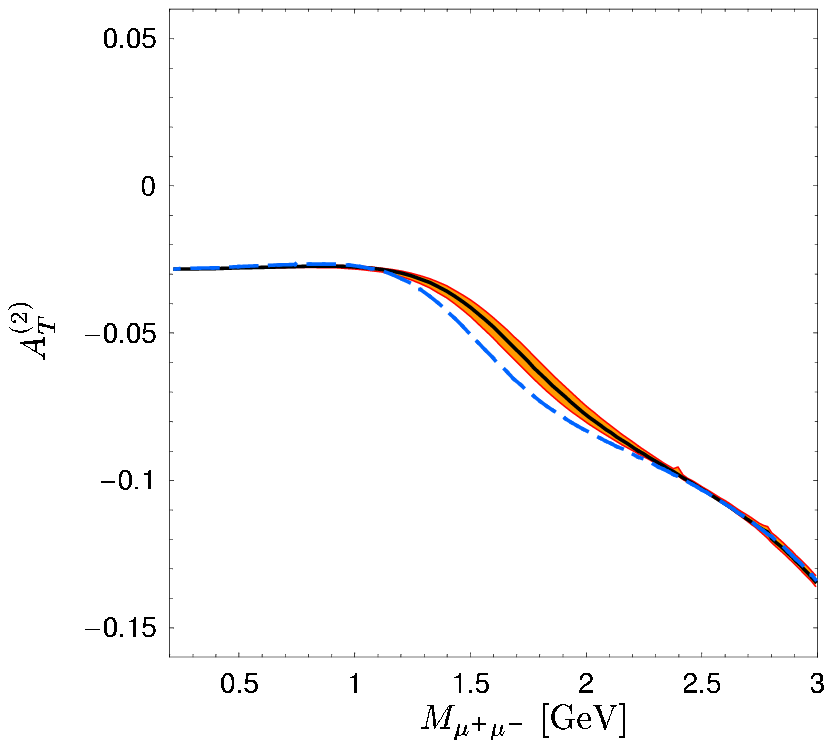}
\includegraphics[scale=0.24]{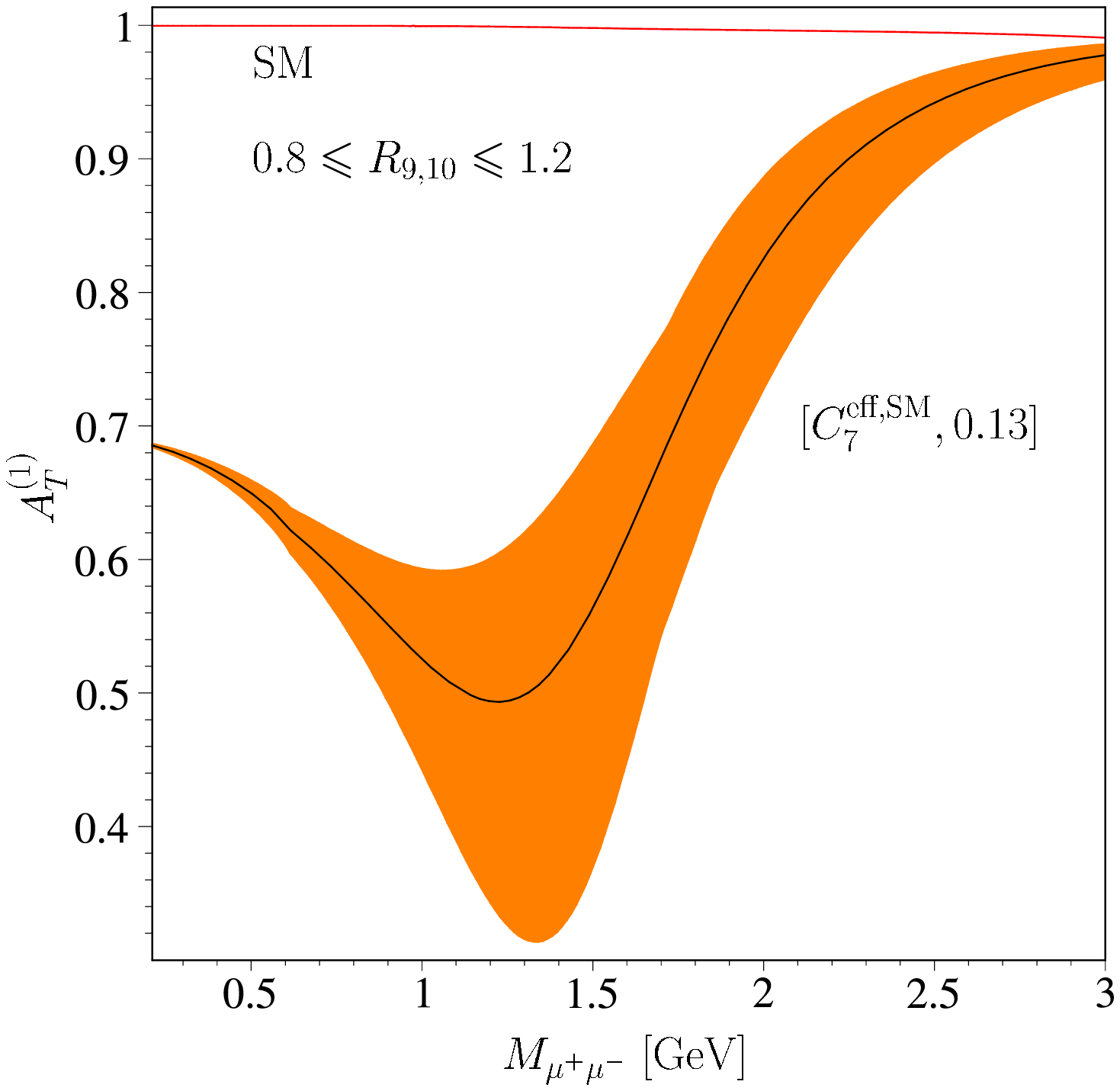}
\includegraphics[scale=0.24]{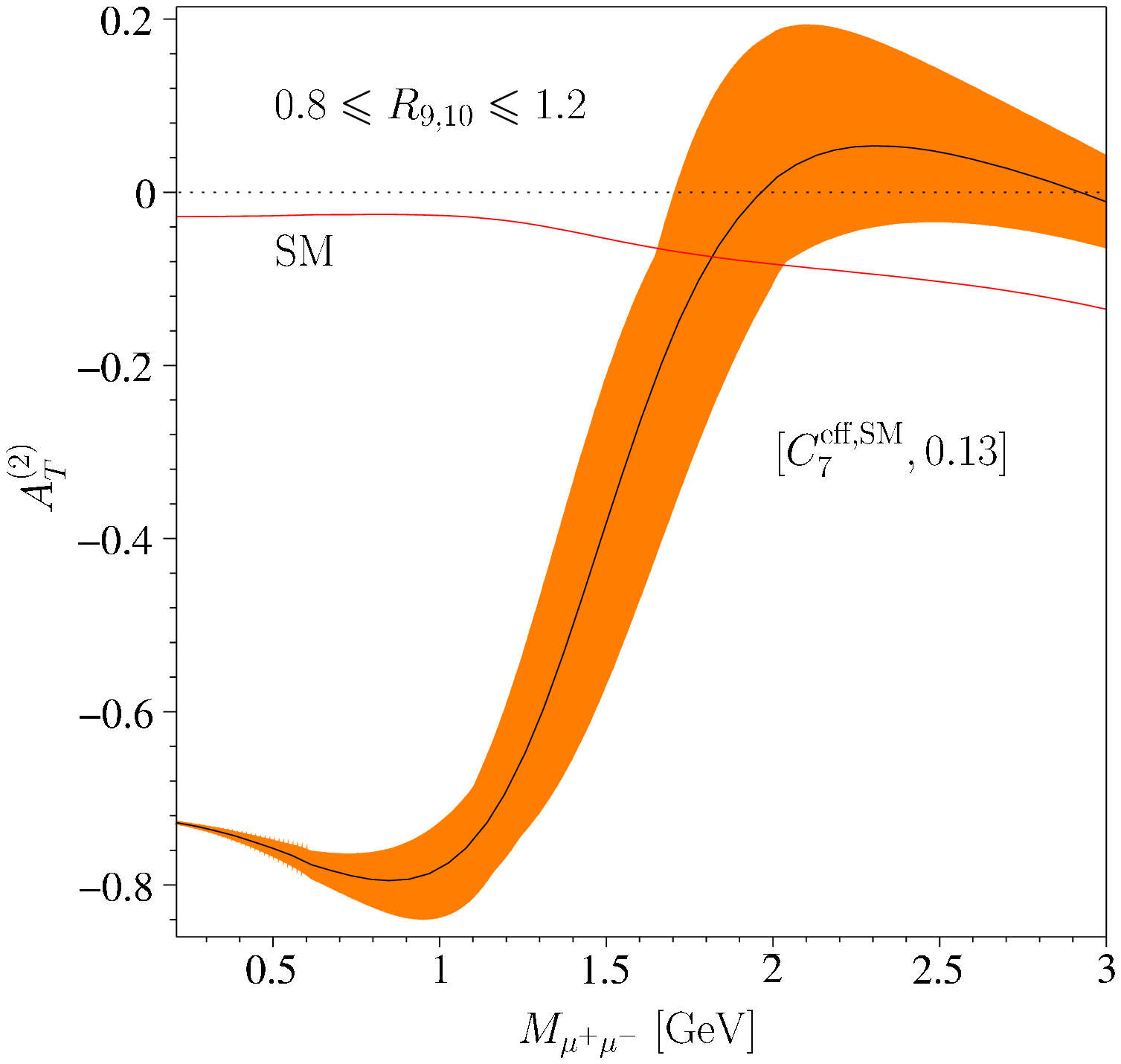}
\caption{First and second: SM predictions for the asymmetries
$A_T^{(1)}$ and $A_T^{(2)}$ as a function of the dimuon mass at LL
(dashed line) and NLL (solid line) including errors (shaded area).
Third and forth: NP impact of RH currents in $A_T^{(1)}$ and
$A_T^{(2)}$ allowing for a NP contribution into $C_{9,10}$ up to
20 \%
as described in \cite{our}}\label{SM:AT1:AT2}
\end{center}
\end{figure}

\end{document}